\documentclass{emulateapj}
\begin{document} 

\title{AC~114: a cluster with a soft X-ray tail}
\shorttitle{AC~114: a cluster with a soft X-ray tail}
\shortauthors{De Filippis E. et al.}

\author{E. De Filippis\altaffilmark{1}}
\author{Mark~W.Bautz\altaffilmark{1}}
\author{M. Sereno\altaffilmark{2,3}}
\author{Gordon~P. Garmire\altaffilmark{4}}

\altaffiltext{1}{Center for Space Research,
               Massachusetts Institute of Technology,
               70 Vassar Street, Building 37,
               Cambridge, MA 02139, USA; bdf@space.mit.edu, mwb@space.mit.edu}

\altaffiltext{2}{INAF - Osservatorio Astronomico di Capodimonte,
		Salita Moiariello, 16
		80131 Napoli, Italia}

\altaffiltext{3}{INFN  and
                Universit\`{a} degli Studi di Napoli ``Federico II'',
                Via Cinthia, Compl. Univ. Monte S. Angelo,
                80126 Napoli, Italia; Mauro.Sereno@na.infn.it}

\altaffiltext{4}{Department of Astronomy and Astrophysics, 
		The Pennsylvania State University, 
		University Park, PA 16802, USA; garmire@astro.psu.edu}

\begin{abstract}
We present {\it Chandra} observations of the galaxy cluster AC~114. The cluster shows a strongly irregular morphology, with signs of multiple merging activity. We report the discovery of a soft X-ray filament originating close to the core of the cluster, curving to the south east for approximately $1.5 \arcmin$. We confirm that X-ray emission is associated with two of three mass concentrations identified in previous gravitational lensing studies of this object~\citep{Nat98,Cam01}. These two mass concentrations are located at opposite ends of the soft filament, and we interpret this as evidence for interaction between them. The northern part of the cluster reveals signs of further dynamical activity. Two sharp discontinuities are detected in both the surface brightness and temperature profiles, evincing another, more recent merger event which took place in, or close, to the cluster core. A preliminary combined mass and lensing analysis shows, in spite of the merger activity, remarkably good agreement between lensing and X-ray masses computed with the assumption of hydrostatic equilibrium.

\end{abstract}


\section{Introduction}
AC~114 (Abell~S1077) is a well-studied galaxy cluster~\citep{Cou87}; at a redshift ${\rm z}~=~0.313$, it has a relatively large velocity-dispersion $\sigma~=~1660\ {\rm km}\ {\rm s}^{-1}$~\citep{Mah01} and a compact core dominated by a cD galaxy. Its strong lensing power was first noticed by~\cite{Sma91} in a survey for bright gravitational lensing arcs. Subsequently, several bright arcs and multiply imaged sources~\citep{Sma95,Nat98,Cam01} were discovered.\\
In this paper we analyze the first {\it Chandra} observation of AC~114. This hot ($kT=8.0\ {\rm keV}$) cluster shows an irregular morphology with a highly elliptical shape. The central cD galaxy, slightly displaced with respect to the centroid of the X-ray emission, is aligned in the general direction of the X-ray brightness elongation. \\
Throughout this paper we quote errors at the $90\%$ confidence level and, unless otherwise stated, we use $H_0=72\ \mathrm{km\ s}^{-1} \mathrm{Mpc}^{-1}$ ($\Omega_m=0.3$, $\Omega_{\Lambda}=0.7$). This implies a linear scale of $4.46\ {\rm kpc}$ per arcsec at the cluster distance $z=0.313$.\\

\section{Observation and Data Analysis}
\label{sec:data_analysis}
The galaxy cluster AC~114 was observed on 2000 November 17 by the {\it Chandra} X-ray Observatory, using the Advanced CCD Imaging Spectrometer (ACIS) (Observation-ID 1562) for 74.3 ksec; the cluster emission lies mainly in the back-illuminated CCD S3.\\
The Chandra Interactive Analysis of Observation (CIAO 2.3) software was used throughout. Level 2 event files supplied by the Chandra X-ray Center (CXC) were updated with new gain maps; bad pixels were removed using the observation-specific bad pixel list provided by the (CXC). Periods of high background rate were eliminated by visual inspection of the light curve in the energy range $10.0-12.0\ {\rm keV}$;  time intervals with rates higher than $70\ {\rm cts}/ \left(100\ {\rm s}\right)$ were eliminated, leading to a final useful exposure time of $71.7$ ksec. \\
Throughout the spectral analysis the blank field background sets (\textsf{acis*sD2000-01-29bkgrndN000*.fits}) by~\cite{Mar03} available in the Chandra Calibration Database (CALDB) were used, after projecting them onto our observation. The blank fields were taken at different locations and in different periods, so the following two  corrections were made:
\begin{itemize}
\item To check for temporal changes in the particle background rate, we compared the count-rates in the cluster and blank fields in the $10.0-12.0\: {\rm keV}$ energy band, where negligible cluster emission is expected. We found a count-rate ratio of $1.015$, which, though remarkably close to unity, was nevertheless used to scale the blank field data.
\item We then looked for a possible residual background due to the soft (${\rm E}<2.0\ {\rm keV}$) Galactic component of the cosmic X-ray diffuse emission which varies with position in the sky. We extracted spectra from the AC~114 field and from the blank fields inside a region where the lowest emission from the cluster was expected and no other cosmic source was observed. In this region (at $\sim 0.8\ {\rm Mpc}$ from the cluster center) a count-rate ${\rm c/r}_{0.3-7.0}=0.179\ {\rm cts\ s}^{-1}$ was measured in this region, very similar to the one measured in the blank fields (${\rm c/r}_{0.3-7.0}=0.174\ {\rm cts\ s}^{-1}$), confirming that the X-ray emission in this area is mainly due to background. We found an excess flux ($\sim 10 \%$) in the cluster field, relative to the background field, at energies less than $2\:{\rm keV}$. The difference between the two spectra was appropriately scaled and then subtracted from each source spectrum throughout the spectral analysis. 
\end{itemize}

\section{Cluster Morphology}
\label{sec:morph_an}
\begin{figure}
\centering
\plotone{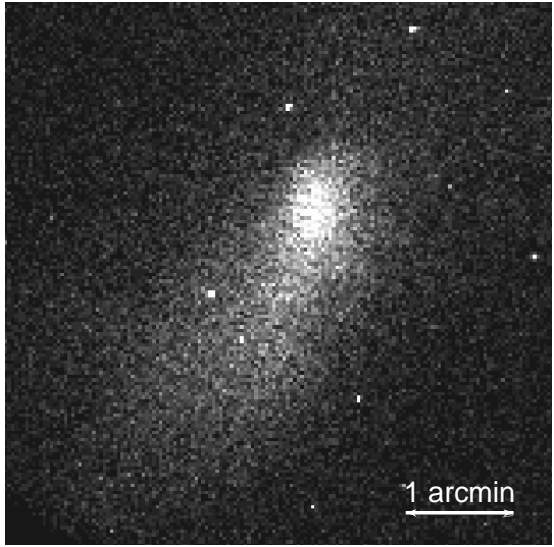}
\caption{\protect{{\it Chandra}} image of AC~114 in the $0.3-7.0\ {\rm keV}$ energy band, binned to $2\arcsec$ pixels.}
\label{fig:03_7_ima}
\end{figure}

\begin{figure}
\centering
\epsscale{1}
\plotone{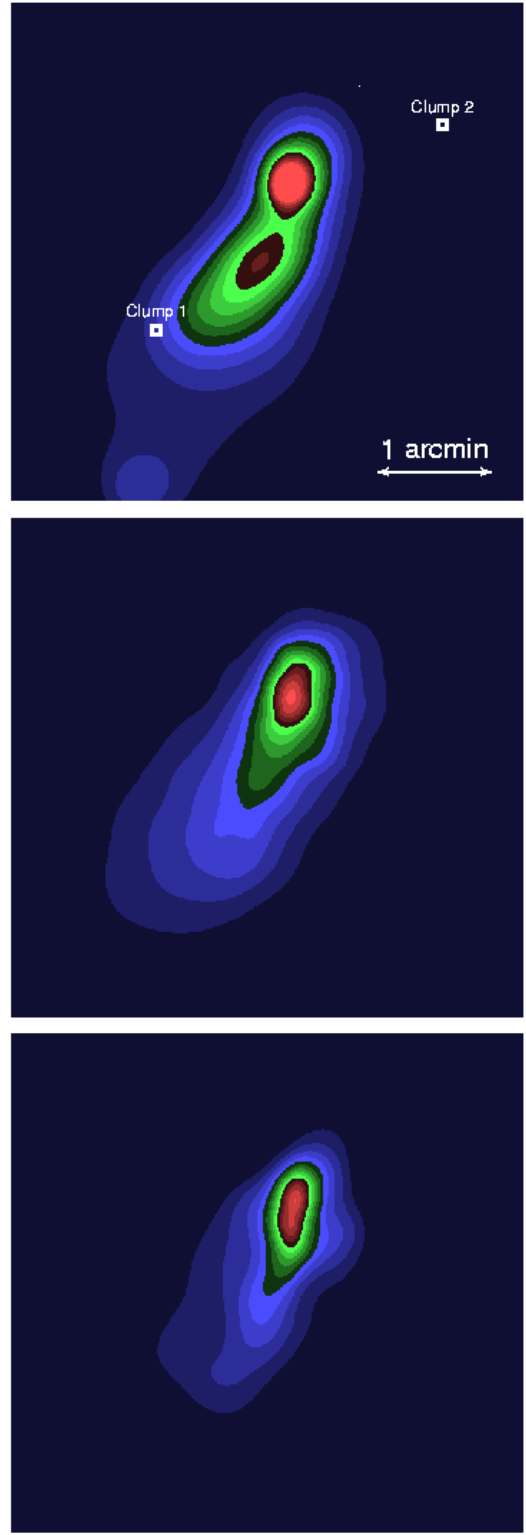}
\caption{Adaptively smoothed, exposure-corrected images of the central 5 arcmin of the cluster in three energy bands: $E<0.3$ keV, $0.3-2.0$ keV and $2.0-10.0$ keV (from top to bottom, respectively). In the top panel {\bf Clump 1} and {\bf Clump 2} show the positions of two additional clumps of matter as estimated with strong and weak lensing measurements (see \S~\ref{sec:lensing}).}
\label{fig:smoo_energy_bands}
\end{figure}

\begin{figure}
\centering
\epsscale{1.0}
\plotone{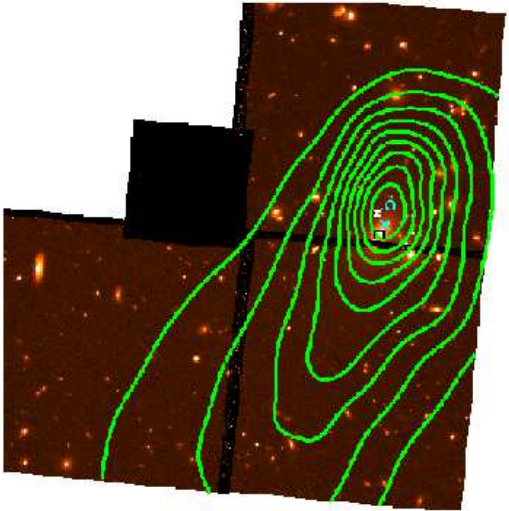}
\caption{HST image of the cluster with $0.3-2.0\ {\rm keV}$ \protect{{\it Chandra}} X-ray contours superposed in green. The positions of the cluster centroid and of the central cD galaxy are plotted as a blue circle and as a black box, respectively. White and blue crosses are at the locations of the X-ray peaks in the ${\rm E} < 0.3\ {\rm keV}$ and $0.3-10.0\ {\rm keV}$ energy bands, respectively.}
\label{fig:HST}
\end{figure}

The {\it Chandra} image of AC~114 in the $0.3-7.0\ {\rm keV}$ energy band is shown in Fig.~\ref{fig:03_7_ima}. The emission arises mainly from the intra-cluster component and from a few additional point sources. While the X-ray surface brightness decreases rapidly out of the cluster center toward the north, and especially the north-east direction (where at just a few arcseconds from the cluster center the X-ray emission decreases sharply almost to the background level), diffuse, elongated fainter emission stretches toward south/south-east down to the edge of the field of view. Since the cluster clearly shows a strongly irregular morphology, we first looked for any energy dependence. We created adaptively smoothed, background and exposure corrected images of the area covered by ACIS detector S3 in three energy bands: $E<0.3\ {\rm keV}$, $0.3-2.0\ {\rm keV}$ and $2.0-10.0\ {\rm keV}$ as shown in Fig.~\ref{fig:smoo_energy_bands} (from top to bottom, respectively). Pixel values of all detected point sources were replaced with values interpolated from the surrounding background regions; the CIAO tools \textsf{wavdetect} and \textsf{dmfilth} were used for this purpose. The images were then adaptively smoothed with minimum and maximum signal-to-noise thresholds set at $3\ \sigma$ and $5\ \sigma$. \\
In the softest energy band ($E<0.3\ {\rm keV}$) the emission is dominated by two extended components: the cluster and an elongated tail-shaped feature extending southeast from the cluster. The elongated southeastern component becomes less prominent at higher energies.\\
The centroid of the overall X-ray emission (computed within a radius of $4\arcmin$ in the $0.3-10.0\ {\rm keV}$ band) is located at $\alpha~=~22^{\rm h}\ 58^{\rm m}\ 48\fs 1$, $\delta=-34\degr\ 47\arcmin\ 59\farcs 4$ (J2000.0), but the cluster does not show a single X-ray peak; noticeable emission is associated with the cluster cD galaxy, which is located about $10\arcsec$ south-east ($22^{\rm h}\ 58^{\rm m}\ 48\fs 4$, $-34\degr\ 48\arcmin\ 08\farcs 5$ J2000.0). The positions of the cluster centroid and of the central cD galaxy are plotted as a blue circle and as a black box, respectively, on the HST image of the cluster in Fig.~\ref{fig:HST}. Superimposed in green are the contours of the X-ray emission in the energy band 0.3-2.0 keV. The offset between the cD galaxy and the centroid of the diffuse intra-cluster medium (ICM) emission is evidence that the latter is most probably not in a state of hydrostatic equilibrium.

\subsection{Two-dimensional model fitting}
\label{sec:morph_fit}
\begin{figure}
\centering
\plotone{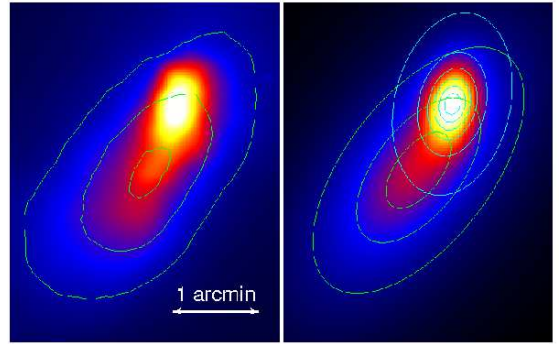}
\caption{{\bf Left panel}: $0.3-10.0\ {\rm keV}$ smoothed image with over-plotted the linearly spaced isointensity contours of the residuals  after subtracting the best-fit model of the northern component from the data. {\bf Right panel}: Image of the best-fitted two-dimensional, double elliptical $\beta$-model of the surface brightness; superimposed are the linearly spaced isointensity contours of the two components of the $\beta$-model.}
\label{fig:comp_fit}
\end{figure}

To parameterize the peculiar morphology of AC~114, we modeled the emission in the central $4\arcmin$ of the cluster in the $0.3-10.0\ {\rm keV}$ band. All point sources in the field were removed. Using the SHERPA software we fitted the cluster surface brightness to several models. Single and double, circular and elliptical, two-dimensional $\beta$-models~\citep{Cav76} have been used. These take the general form: 
\begin{displaymath}
S(x,y)=S_0\left(1+\left(\frac{\sqrt{x_1^2+y_1^2/\epsilon^2}}{r_c}\right)^2\right)^{-3\beta+1/ 2}
\end{displaymath}
with:
\begin{eqnarray}
\nonumber x_1=(x-x_c)\cos\theta+(y-y_c)\sin\theta\\
\nonumber y_1=(y-y_c)\cos\theta-(x-x_c)\sin\theta
\end{eqnarray}
where $r_{c}$ is the core radius, $\beta$ the slope parameter, $\epsilon$ the axial ratio, $\theta$ the position angle (north over east) and $x_c$ and $y_c$ the location of the center of the model. The X-ray emission turns out to be best described by a simultaneous fit of two separate components. The northern component is centered on the centroid of the cluster X-ray emission and has an axial ratio $\epsilon=0.61$. The center of the southern component is located $\sim 0.9\arcmin$ south-east of the X-ray centroid and  is centered on the elongated soft tail visible south of the cluster. This southern component is more elliptical (axial ratio $\epsilon=0.50$) than the northern one, and is rotated about $15 \degr$ with respect to it. The  positions, shapes, orientations and relative amplitudes of these two model components are listed in Table~\ref{tab:2Dfit}. The sum of the components is shown in a false color image in the right panel in Fig.~\ref{fig:comp_fit}; superposed are the isointensity contours of  each component.  To better illustrate the shape of the softer and fainter southern component, we subtracted the best-fit model of the northern component from the data in the $0.3-10.0\ {\rm keV}$ band. Smoothed contours of the residuals are shown in the left panel in Fig.~\ref{fig:comp_fit}, over-plotted on the smoothed data. The regular, elliptical shape of the southern component is evident. For the remainder of  this paper this Southern component will be called the ``Tail''.\\

\tabletypesize{\scriptsize}
\def\arraystretch{1.0}
\begin{deluxetable*}{llccccc}
\tablecolumns{7}
\tablewidth{0pt}
\tablecaption{Fit parameters of the 2D elliptical double-$\beta$ model}
\tablehead{
\colhead{Component}  &\colhead{$x_c,y_c$}  &\colhead{$r_{c}$}   &\colhead{$\beta$} &\colhead{$\epsilon$} &\colhead{$\theta$} & \colhead{A}\\
\colhead{}	   &\colhead{(J2000.0) }   & \colhead{[$\arcmin$]} &\colhead{}	   &	\colhead{}       &\colhead{[$\degr$]}&\colhead{[cts s$^{-1}\ (\arcmin)^{-2}$]}
}
\startdata
Northern    &$\alpha=22^{\rm h}:58^{\rm m}:48\fs 1$  &$1.0\pm0.8$  &$1.7\pm1.6$    &$0.61\pm 0.08$       &$167\pm7$	 &$0.19\pm0.03$\\
	    &$\delta=-34\degr:47\arcmin:59\farcs8$ &	&	&	&	&  \\
Southern    &$\alpha=22^{\rm h}:58^{\rm m}:50\fs 3$  &$1.3\pm1.0$  &$0.6\pm 0.4$    &$0.50\pm0.04$      &$143\pm3$	 &$0.086\pm0.008$\\
 	    &$\delta=-34\degr:48\arcmin:46\fs 1$ &	&	&	&	&  \\
\enddata
\tablecomments{Fit parameters of the 2D elliptical double-$\beta$ model of the cluster: $x_c,y_c$ is the central position of each component, $r_{c}$ the core radius, $\beta$ the slope parameter, $\epsilon$ the axial ratio, $\theta$ the orientation angle (north over east) and A the amplitude of the model at $x_c,y_c$.}
\label{tab:2Dfit}
\end{deluxetable*}


\section{Spectral Analysis}

\subsection{Global Cluster Properties}
\label{sec:spec_an}
A global spectral analysis for the cluster was performed inside a circle centered at $\alpha=22^{\rm h}\ 58^{\rm m}\ 48\fs 1$, $\delta=-34\degr\ 47\arcmin\ 59\farcs 8$ (J2000.0) and within a radius of $2.5\arcmin$; this region includes both the cluster and the tail. The emission from all detected point sources inside this region was excluded. We fitted the data to an absorbed XSPEC isothermal plasma emission code by~\cite{Kaa93} including the FeL calculations of~\cite{Lie95}~\footnote{MeKaL code hereafter.}, folded through the appropriate response matrices, corrected for the ACIS time dependent absorption due to molecular contamination. An excess of soft emission is observed in the spectrum at energies below 0.5 keV, most probably due to the emission of the soft Tail. We were able to obtain a good spectral fit only when energies below 0.5 keV were excluded. 
(We return to this point in Section~\ref{sec:spec_NeS} below.) 
Results obtained in this spectral analysis are summarized in Table~\ref{tab:spec_cluster}.
For the spectral range ${\rm E}>0.5\ {\rm keV}$, results are in agreement with previous findings by~\cite{All00}.\\
The hydrogen column density along the line of sight is not well constrained by our fit, and may be subject to systematic errors in the calibration of the (changing) low-energy response of ACIS.  We therefore analyzed the cluster spectrum in the restricted energy range $1.5-7.0\ {\rm keV}$, where the value of ${\rm n_H}$ does not affect the fit. Results, shown in Table~\ref{tab:spec_cluster}, are consistent, within the uncertainties, with the fits over the $0.5 - 7.0$ keV band. We conclude that spectral fits over the broader band are reliable.\\

\tabletypesize{\scriptsize}
\def\arraystretch{1.0}
\begin{deluxetable*}{lcccccc}
\tablecolumns{8}
\tablewidth{0pt}
\tablecaption{MeKaL spectral fits for the cluster central $2.5\arcmin$ and for two ellitpical regions, one centered on the Tail (region A) and the second one symmetric respect to the cluster centroid (region B).}
\tablehead{
\colhead{Region}&\colhead{Energy}& \colhead{kT}& \colhead{Metallicity}& \colhead{Redshift} & \colhead{HI column density} & \colhead{$\chi^2/{\rm d.o.f.}$} \\
\colhead{}&\colhead{[keV]}& \colhead{[keV]}&\colhead{[${\rm Z}_\sun$]}& \colhead{} & \colhead{$\left [10^{20} {\rm cm}^{-2}\right]$}&\colhead{}
}
\startdata
Cl. center & $0.5-7.0$& $8.0\pm0.5$ & $0.25\pm 0.09$&  $0.313$ (fixed) &$0\pm2$ & $292/271$\\
Cl. center & $0.5-7.0$& $8.0\pm0.5$ & $0.30\pm 0.09$&  $0.34^{+0.01}_{-0.02}$ & $0\pm2$ &$282/270$\\
Cl. center & $0.5-7.0$& $11.0^{+1.0}_{-0.8}$ & $0.20\pm 0.12$&  $0.313$ (fixed) & $1.31$ (fixed) &$343/272$\\
Cl. center & $1.5-7.0$& $8.6^{+1.2}_{-0.9}$ & $0.24\pm 0.10$&  $0.313$ (fixed) & $7\pm10$ & $177/203$\\
A	& $0.1-7.0$& $7.5^{+1.8}_{-0.9}$ & $0.37\pm 0.20$&  $0.313$ (fixed) &$4.1\pm0.6$ & $187/93$\\
A	& $0.1-7.0$& $12.2^{+2.5}_{-2.2}$ & $0.37\pm 0.33$&  $0.313$ (fixed) &$1.31$ (fixed) & $221/94$\\
B	& $0.1-7.0$& $13^{+10}_{-5}$ & $0.4^{+0.8}_{-0.4}$&  $0.313$ (fixed) & $3.9\pm1.2$ &$31/39$\\
B	& $0.1-7.0$& $26^{+25}_{-10}$ & $0.4^{+1.1}_{-0.4}$&  $0.313$ (fixed) & $1.31$ (fixed) &$39/40$\\
\enddata
\label{tab:spec_cluster}
\end{deluxetable*}

\subsection{Flux and Luminosity}
The cluster emission can be traced out to a radius of $3.5 \arcmin\sim 940\ {\rm kpc}$. Within this radius the cluster bolometric flux and luminosity (at the measured temperature of ${\rm kT}=8.0\pm0.5\ {\rm keV}$) are:  $f_{\rm X}{\rm (bol)}=5.6^{+0.3}_{-0.2}\times 10^{-12}\ {\rm erg\ s}^{-1}\ {\rm cm}^{-2}$, $L_{\rm X}{\rm (bol)}=12.9^{+0.7}_{-0.3}\times 10^{44}\ {\rm erg\ s}^{-1}$.\\
AC~114 follows the typical L$_X$-T cluster relation, with an expected $L_{\rm X}^{\rm exp}{\rm (bol)}=12.7\pm 2.6 \times 10^{44}\ {\rm erg\ s}^{-1}$~\citep{Arn99} in agreement with our findings. 
In contrast, we find that AC~114 does not fall on established trends relating X-ray properties to velocity dispersion. If we adopt the higher velocity dispersion, $\sigma=1660^{+128}_{-106}\ {\rm km}\ {\rm s}^{-1}$, computed by~\cite{Mah01} within a radius of $1\ {\rm Mpc}$, bolometric luminosities and temperatures as high as $L_{\rm X}^{\rm exp}{\rm (bol)}=45.6^{+2.4}_{-2.2}\times 10^{44}\ {\rm erg\ s}^{-1}$ and ${\rm kT}^{\rm exp}=13\pm5\ {\rm keV}$ are predicted~\citep{Mah01,Wu99}. Even the lower velocity dispersion ($\sigma=1380^{+128}_{-71}\ {\rm km}\ {\rm s}^{-1}$, computed within $0.9\ {\rm Mpc}$) reported by~\cite{Gir01} leads to an expected value of $L_{\rm X}^{\rm exp}{\rm (bol)}=19.9^{+2.4}_{-2.2}\times 10^{44}\ {\rm erg\ s}^{-1}$, significantly higher than the value we measure; the expected gas temperature of ${\rm kT}^{\rm exp}=10\pm4$ keV is instead consistent to the measured value.\\
The mean $L_{\rm X}-\sigma$ and $\sigma-{\rm T}$ relations presumably characterize dynamically relaxed clusters, so departures from these relations may be expected for unrelaxed systems~\citep{Wu99}. Our results are thus consistent with the view that AC~114 is not fully relaxed. While the significant discrepancy between the published velocity dispersions might suggest that this conclusion tentative at best, may also, in itself, be indicative of the dynamical complexity of the cluster. 

\subsection{The Tail}
\label{sec:spec_NeS}
\begin{figure}
\centering
\plotone{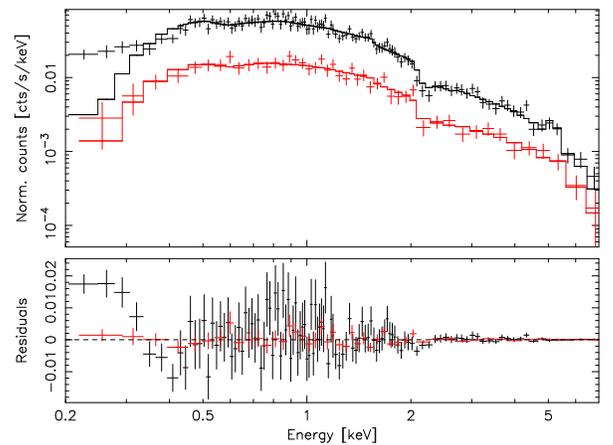}
\caption{Spectra, with superimposed  best-fit models (top panel), and residuals (bottom panel), extracted from the region in the south where the Tail is located (black line), and from a region in the north symmetric respect to the cluster center (red line).}
\label{fig:soft_excess}
\end{figure}

In order to characterize the spectral properties of the elongated Tail, we computed and then compared spectra extracted within two elliptical regions: one centered on the Tail and the second located at $22^{\rm h}\ 58^{\rm m}\ 44\fs 0$, $-34\degr\ 47\arcmin\ 10\farcs 1$, diametrically opposite the Tail with respect to the cluster center. Both regions have shapes and orientations given by the Southern component in Table~\ref{tab:2Dfit}; the semi-major axis is equal to $0.5 \arcmin$ for each region. The two resulting spectra, with the respective best-fit MeKaL models superimposed, are shown in Fig.~\ref{fig:soft_excess}; the spectrum extracted from the Tail is plotted as a black line, while the one from the region in the north is shown in red. The MeKaL model was fitted in the $0.5-7.0\ {\rm keV}$ energy band. Resulting best-fit parameters for both regions are consistent with the global cluster spectrum (see Table~\ref{tab:spec_cluster}).\\
Although the two regions are symmetric respect to the cluster centroid, their emission spectra differ significantly (see Table~\ref{tab:soft_excess} and Fig.~\ref{fig:soft_excess}). First, the Tail is brighter by a factor of three. Second, the Tail shows a strong excess of emission with respect to the fitted model at energies below $0.4\ {\rm keV}$; no such soft excess is present in the northern region.\\
In the Tail region, we have a superposition of the emission from the Tail and from the cluster. At energies above $1.5\ {\rm keV}$, the spectra of the two regions are similar, but the Tail is brighter. This result suggests the Tail region contains plasma with a temperature similar to that of the Northern region. It is thus natural to conclude that the ICM in these two regions are dynamically related, and that the cluster emission is more elongated in the southern direction, possibly due to some recent dynamic interaction. \\
In order to characterize the excess soft emission in the southern region, we subtracted the two spectra extracted above, after having rescaled the northern component to have the same flux as the southern component at energies higher than $1.5 \ {\rm keV}$. The limited number of photon counts, and  calibration uncertainties in the very low energy range, make a precise characterization of the spectral properties of the Tail rather difficult. An estimate of its temperature, however,  has been obtained from broadband colors. Color ratios among three energy bands ($0.1-0.8\ {\rm keV}$, $0.8-1.5\ {\rm keV}$ and $1.5-8.0\ {\rm keV}$) 
were computed for simulated spectra as a function of temperature. We find a direct correspondence between the color ratio ${\rm CR}={\rm counts}[(0.8-1.5)\ {\rm keV}]/{\rm counts}[(1.5-8.0)\ {\rm keV}]$ and the temperature~\footnote{$kT={\rm EXP} \left( 1.003\pm 0.003 - 2.48\pm0.04 \times \log({\rm CR}) \right)$.}. The observed ${\rm CR}=2.97$ implies a temperature for the Tail of $kT=0.7\ {\rm keV}$ and a bolometric luminosity, within $0.5\arcmin$ of $2.7\times 10^{43}\ {\rm erg\ s}^{-1}$. We stress that the given values of the temperature and of the bolometric luminosity of the Tail are only tentative estimates since an exact spectral characterization of the region  could not be obtained. Furthermore normalizing the spectra extracted in the two different regions and creating count-rate ratios using two energy ranges can be affected by systematic errors depending on uncertainties in the background estimate which can affect the Northern and the Southern regions in different ways and on the choice of the threshold temperatures.\\
A soft filament of similar shape, but of smaller dimensions ($\sim 80\ {\rm kpc}$) was observed in the galaxy cluster Abell~1795~\citep{Fab01} and explained by the authors as a possible cooling wake, caused by the motion of the central cD galaxy. In the case of AC~114 the high bolometric luminosity of the Tail, and the fact that it extends for $\sim 400\ {\rm kpc}$, tends to exclude the motion of the cD galaxy as a possible cause of its origin. An infalling group, stripped of its intra-group medium when it passed close to the cluster core is a more likely explanation for the soft component in this cluster.\\
Spatially dependent degradation in the quantum efficiency and calibration uncertainties could in principle be affecting our imaging and spectroscopic results at such low energies. Our detection of a soft excess is spatially confined in the region of the Tail; significant sensitivity fluctuations at that specific position of the instrument would be required to cause such an effect, which should therefore be detectable also in other observations. We have analyzed spatial photon distribution in a low (${\rm E} < 0.3\ {\rm keV}$) and in a broad energy range ($0.3-7.0\ {\rm keV}$) both in the blank fields and in a deep observations of a galaxy cluster. In both cases we did not detect flux increase at the location of the Tail. Spectra extracted from the blank fields within the two elliptical regions (described at the beginning of this section) also show no significant difference at low energies. Calibrations issues and spatially dependent degradation in the instrument quantum efficiency should therefore not be affecting our detection of the soft Tail.\\

\begin{deluxetable}{ccc}
\tabletypesize{\scriptsize}
\tablecolumns{8}
\tablewidth{0pc}
\tablecaption{Count-rates in different energy bands}
\tablehead{
\colhead{Region} & \colhead{c/r ($0.1-0.4\ {\rm keV}$)}& \colhead{c/r ($0.4-8.0\ {\rm keV}$)}\\
\colhead{}       & \colhead{[cts sec$^{-1}$]} 	   & \colhead{[cts sec$^{-1}$]}
}
\startdata
south & $30.9\times 10^{-4}$& $4.0\times 10^{-2}$ \\
north & $7.5\times 10^{-4}$ & $1.5\times 10^{-2}$ \\
\enddata
\label{tab:soft_excess}
\end{deluxetable}


\section{A Recent Merger Close to the Cluster Core?}
\label{sec:cold_front}
\begin{figure*}
\centering
\plotone{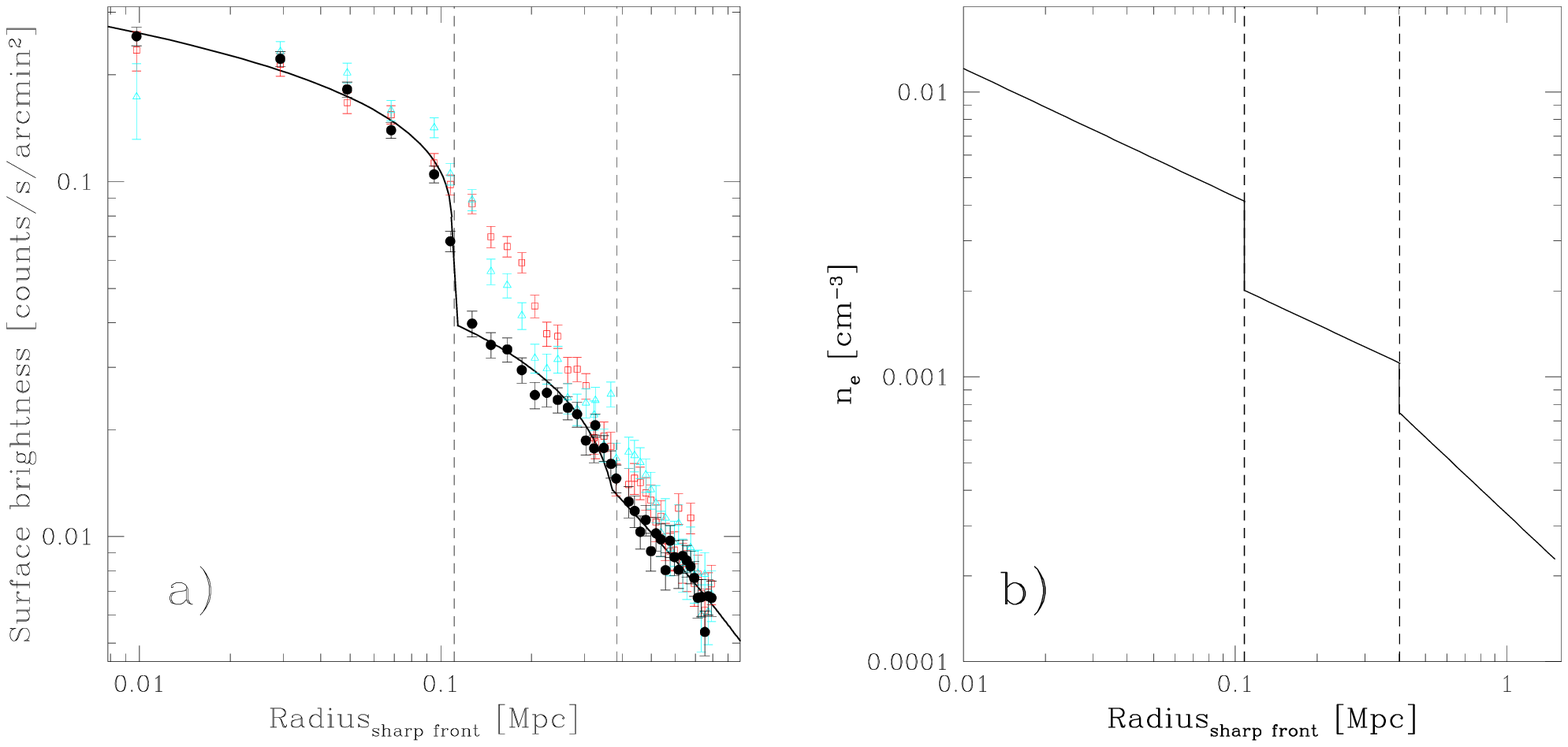}
\plotone{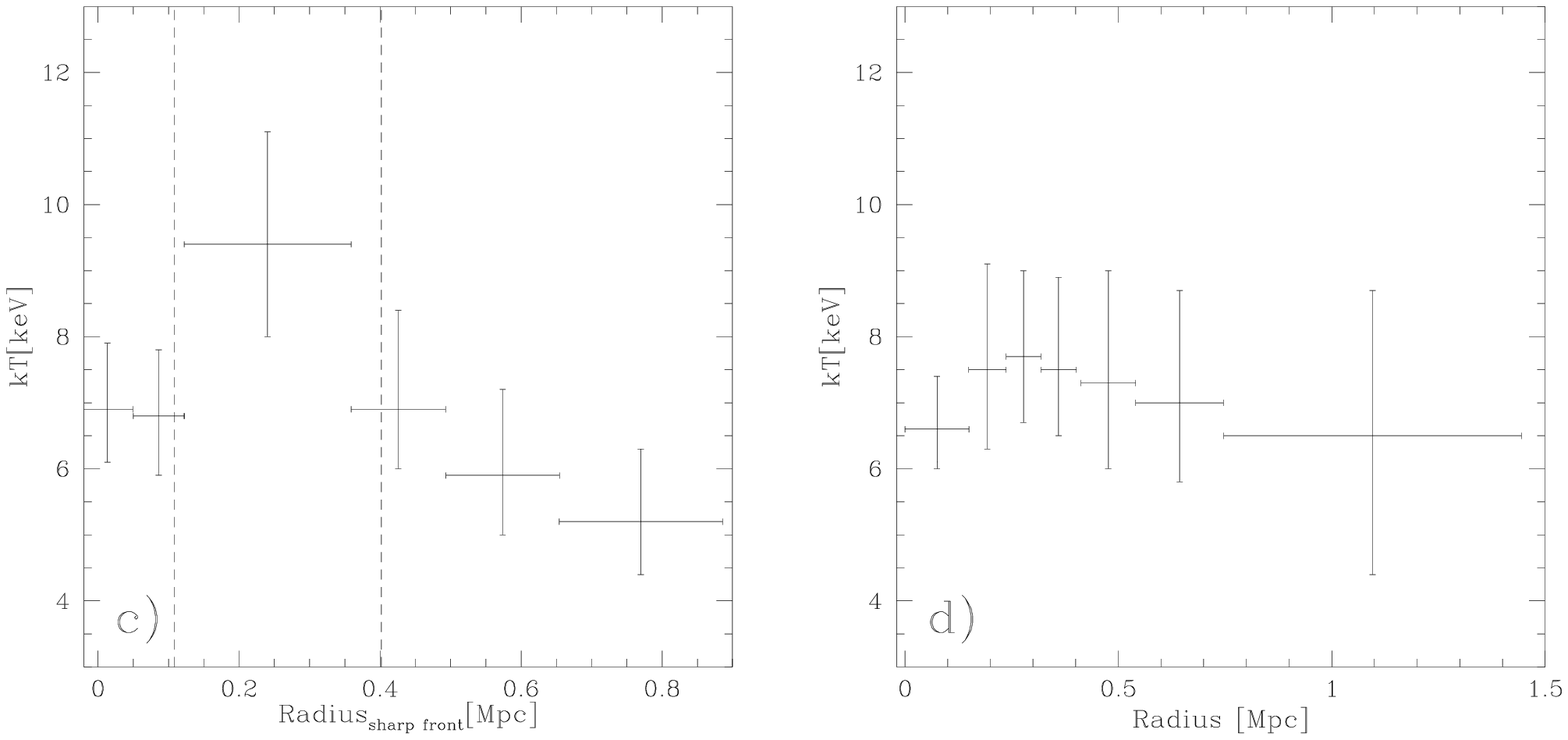}
\plotone{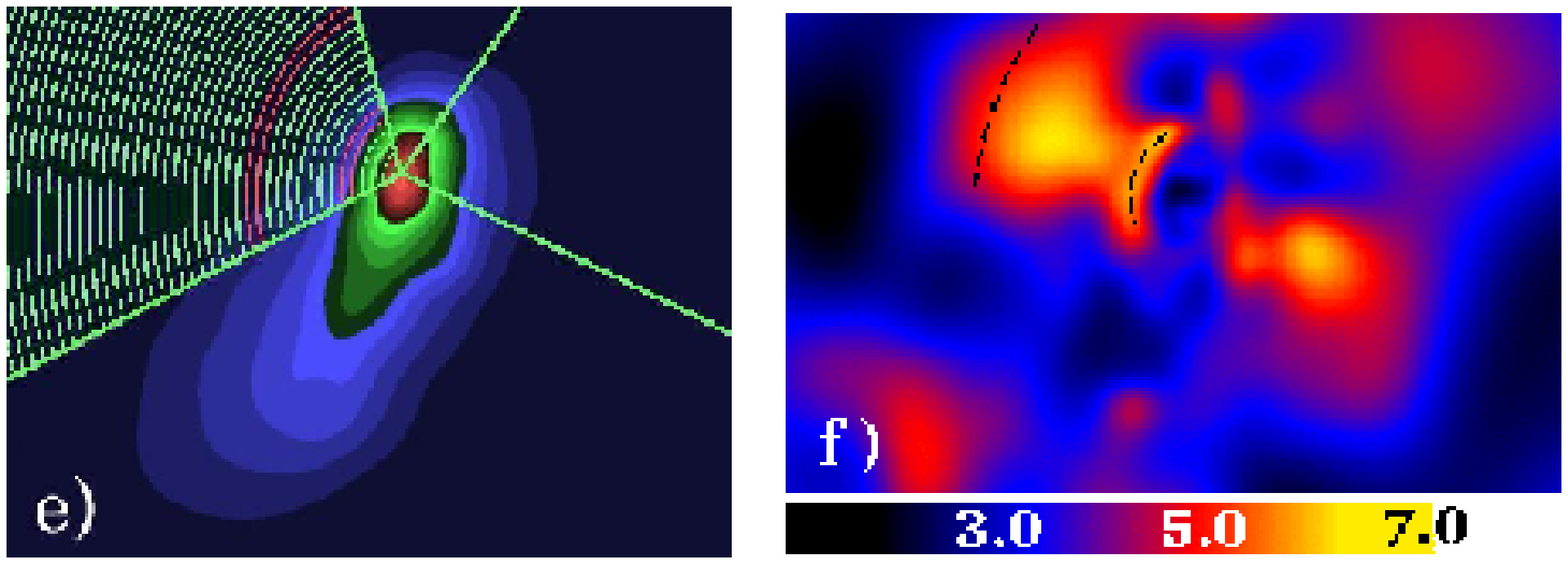}
\caption{{\bf a)}: surface brightness profiles extracted in the north-west, north and north-east wedges (plotted in red, light blue and black, respectively; superimposed as a solid black line is the best-fit projected model of the north-east profile). {\bf b)}: density model. {\bf c)}: temperature profile extracted in the north-west wedge. {\bf d)}: temperature profile extracted in a $150 \degr$ wide wedge (centered at the cluster X-ray centroid) in the southern portion of the cluster. {\bf e)}: adaptively smoothed color image of the cluster in the energy band 0.3-2.0 keV; superposed are the wedges within which the three surface brightness profiles, shown in the top left panel, have been extracted. The pink curves show the position of the fronts ${\rm r}_{12}$ and ${\rm r}_{23}$. {\bf f)}: HR image where colors from blue, to red an yellow indicate increasingly harder ratios, as labeled in the color scale at the bottom of the image. The two fronts ${\rm r}_{12}$ and ${\rm r}_{23}$ are plotted as black dashed black curves.}
\label{fig:cf}
\end{figure*}

Fig.~\ref{fig:03_7_ima} shows a sharp cutoff in the cluster brightness toward the north/north-east, while the cluster emission is more diffuse and extended in the other directions suggesting that the cluster, or an internal substructure, is moving toward the north/north-east direction creating a sharp edge in the front and a more diffuse feature behind.\\
To explore this possibility, we extracted surface brightness profiles from three contiguous wedges in the northern portion of the cluster. The wedges are plotted on panel e) of Fig.~\ref{fig:cf}, superposed on the $0.3-2.0\ {\rm keV}$ smoothed image of the cluster; the annuli are centered at the cluster X-ray centroid and have an axial ratio ${\rm b}/{\rm a}=0.8$. The resulting surface brightness profiles, plotted in red, light blue and black in panel a) of Fig.~\ref{fig:cf}, are extracted from the north-west, north and north-east wedges, respectively. Although all three profiles show a steep decrease as a function of radius, the north-east profile reveals two discontinuities: a sudden decrease of a factor of $\approx 4$ at a distance of around $0.1$ Mpc from the cluster center, and a second smaller decrease between $0.3$ and $0.4$ Mpc.\\
In the area between these two jumps the temperature of the ICM shows a maximum (panel c) in Fig.~\ref{fig:cf}). Although the errors in the temperature profile are large, the temperature increase can be clearly seen as the sharp hard curved front in the hardness ratio (HR) map (panel f) of Fig.~\ref{fig:cf}); the two dashed black lines show the positions where the two discontinuities, both in surface brightness and temperature, are observed. A second hard region is observed on the western side of the cluster core, but neither a sharp edge in the HR map, nor drastic jumps in the surface brightness and temperature, are observed there. The HR map has been obtained from the ratio of the adaptively smoothed images in the $1.0-10.0\ {\rm keV}$ band and below $0.5\ {\rm keV}$, with point sources excluded; the use of a smoothing technique with a fixed sigma does not change features in the final HR map. In the resulting image, colors from dark blue to red and yellow, correspond to increasingly harder ratios.\\
Similar discontinuities in the surface brightness, with corresponding jumps in the gas temperature, have been observed in other clusters~\citep{Vik01,Mar02}. Also, recent adiabatic hydro-dynamical simulations by~\cite{Mat03} show how, during a cluster merger, two shock fronts form, and subsequently expand from the central regions into the ICM, right after the cores of two merging structures pass one another; AC~114 shows very similar features, with two hard regions located on opposite sides of the X-ray centroid. While the western hard region is not as well defined, in the eastern one we have seen how a sharp front with a corresponding temperature increase is clearly visible. The two discontinuities in the surface brightness emission of the cluster are observed in projection on the plane of the sky. To obtain an estimate of the positions and amplitudes of the two jumps, we approximate the density distribution in the cluster with an analytical model. We use a model composed of three power laws centered in the cluster centroid, having different amplitudes and slopes in three regions: interior to the inner discontinuity, between the two discontinuities, and exterior to the outer discontinuity.
We then fit the X-ray surface brightness distribution predicted by this density model to the observed profile. The free parameters are the three power law indexes, the two radii and the two amplitudes of the jumps ($\alpha_1$, $\alpha_2$, $\alpha_3$, r$_{12}$, r$_{23}$, ${\rm A}_1$ and ${\rm A}_2$ respectively). The best-fit surface-brightness model, which provides a reasonably good fit to the data, and the associated (3D) density model, are shown in panels a) and b) of Fig.~\ref{fig:cf}. The best-fit radial slopes are: $\alpha_1=0.90\pm 0.19$, $\alpha_2=0.89\pm 0.25$ and $\alpha_3=1.80\pm 0.15$ and the positions of the two jumps are at distances ${\rm r}_{12}=0.108\ {\rm Mpc}$ and  ${\rm r}_{23}=0.401\ {\rm Mpc}$ from the cluster center. The amplitude of the jumps in the gas density are easily computed from the square root of the relative amplitudes of the three best-fit surface brightness models:
\begin{displaymath}
\frac{\rho_1}{\rho_2}=2.1\pm0.5;\ \ \ \ \frac{\rho_2}{\rho_3}=1.49\pm0.16
\end{displaymath}
At ${\rm r}_{12}$ we observe a sharp gas density decrease and an increase in the gas temperature which, together with the only small decrease observed in the pressure, indicates that ${\rm r}_{12}$  does not correspond to a shock front.\\
At ${\rm r}_{23}$ we instead observe a sharp decrease of both the gas density and the temperature. This is consistent with the physical conditions expected if the gas was experiencing a shock.\\
AC~114 could therefore be going through a merging process in its core, with consequences similar to what is observed in clusters such as Abell 2142~\citep{Mar00}.  Within ${\rm r}_{12}$ we would then be  observing the core of one of the two sub-clusters, moving toward the northeast, after a relatively recent merger. If this sub-cluster is moving supersonically, a bow shock would form in front of it, with a subsequent sudden decrease both in surface brightness and temperature, as we observe at ${\rm r}_{23}$. \\
Given the relatively large errors in the temperature and in the fit of the projected gas density model, it is impossible to draw precise quantitative conclusions. We have, nevertheless, made crude estimates of the shock parameters. From the density jump across the bow shock at ${\rm r}_{23}$ we estimate the Mach number of the core to be ${\rm M}=1.3\pm 0.1$. This should correspond to a temperature and pressure ratio across ${\rm r}_{23}$ 
of $\Delta kT_{23}^{\rm est}= 1.3\pm 0.2 \ {\rm keV}$ and $\Delta P_{23}^{\rm est}=1.8\pm 0.4\ {\rm erg}\ {\rm cm}^{-3}$. The observed temperature and pressure discontinuities across ${\rm r}_{23}$ are $\Delta kT_{23}^{\rm obs}= 1.4^{+0.3}_{-0.2} \ {\rm keV}$ and $\Delta P_{23}^{\rm obs}=2.0^{+0.4}_{-0.3} {\rm erg}\ {\rm cm}^{-3}$ which are consistent with the  estimated Mach number.\\
A second method for computing the Mach number due to~\cite{Moe49} and applied to clusters by~\cite{Vik01} yields a similar estimate: ${\rm M}=1.15$.\\


\section{Mass Distribution}
\label{sec:mass_analysis}
\begin{figure}
\centering
\plotone{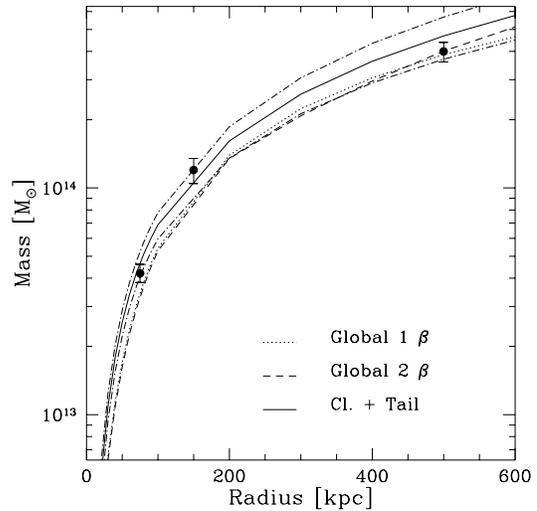}
\caption{Mass profiles computed assuming the \textsf{Global}-1$\beta$-model, \textsf{Global}-2$\beta$-model and the \textsf{N+S} model, plotted as a dotted, dashed and solid line, respectively. The $90 \%$ confidence levels for the \textsf{N+S} model are plotted as dot-dashed lines. Results from lensing from~\cite{Nat98} are also plotted.}
\label{fig:massa}
\end{figure}

The irregular morphology and the apparently unrelaxed dynamical state of AC114 suggest that the ICM may not be in hydrostatic equilibrium. We therefore estimate the mass profile in three different ways. We begin with the simple assumption that all the observed hot gas is part of a single structure (\textsf{Global} model), first modeling its surface brightness distribution with a single $\beta$-model and subsequently with a double $\beta$-model. Finally, we consider the hypothesis that the cluster is composed of two dynamically independent systems -the Northern component and the Southern Tail- (\textsf{N+S} model).\\
For the first step of our mass analysis, we extracted 1D surface brightness profiles within elliptical concentric annuli spaced by 5 arc-seconds, with ellipticity, position angle and center matching those of the northern component (see Table~\ref{tab:2Dfit}). The distance $r$ from the cluster center is measured along the major axis of the ellipses. The profiles were extracted in the 0.3-7.0 keV energy range. We extracted background profiles from the blank field data, in the same regions and energy band, scaled as described in Section~\ref{sec:data_analysis} above. Corresponding profiles extracted from the exposure maps provided the required average values of the effective exposure time within each annulus. The best-fit single $\beta$-model to the resulting surface brightness profile is described by: $S_0=0.253\pm0.016\ [{\rm cts}\ {\rm s}^{-1}\ {\rm arcmin}^{-2}]$, $\beta=0.423\pm0.007$, $r_c=0.37\pm0.03\ {\rm arcmin}$, $\chi^2_{\rm red.}=4.3$. A double $\beta$-model increases the quality of the fit, providing an accurate description of the cluster profile despite its strong irregularity: $S_{01}=0.213\pm0.016\ [{\rm cts}\ {\rm s}^{-1}\ {\rm arcmin}^{-2}]$, $\beta_1=0.9\pm0.5$, $r_{c1}=0.557\pm0.045\ {\rm arcmin}$, $S_{02}=0.056\pm0.008\ [{\rm cts}\ {\rm s}^{-1}\ {\rm arcmin}^{-2}]$, $\beta_2=0.60\pm0.06$, $r_{c2}=1.8\pm0.4\ {\rm arcmin}$, $\chi^2_{\rm red.}=1.1$.\\
The electron density $n_{\rm e}$ is then obtained by inverting the model which describes the surface brightness distribution. In the spherically symmetric case the inversion of the double $\beta$ model has been resolved by~\cite{Xue00}. Since we are dealing with elliptical distributions, we had to extend the formalism to a more general ellipsoidal symmetry; we considered an oblate ellipsoid with the $z$-axis along the line of sight, and we did not consider a possible inclination of the principal axes. Under these assumptions, the electronic density for the single $\beta$-model reads
\begin{displaymath}
n_{\rm e}^{1\beta} (m) = n_{\rm e0} \left( 1+\frac{m^2}{r_{\rm c}^2}\right)^{-\frac{3\beta}{2} }
\end{displaymath}
where $ m \equiv \sqrt{ x_1^2+\frac{y_1^2}{\epsilon^2}+z^2}$ defines an ellipsoid coordinate. For a double concentric $\beta$-model
\begin{displaymath}
n_{\rm e}^{2\beta} (m) = \left\{ \sum_{i=1}^2 \left[ n_{\rm e}^{1\beta} \left( m;n_{{\rm e0}i}, r_{{\rm c}i},\beta_i \right) \right]^2 \right\}^{1/2}
\end{displaymath}
At large radii from the cluster center ($\sim r_{500}$: radius corresponding to an overdensity $\Delta=500$ with respect to the critical density) an assumption of a prolate geometry leads to negligible differences in the cluster total mass and gas mass fraction; differences in the total mass of clusters, even in compressed or elongated geometries, have been estimated to be less than $4\%$~\citep{Pif03}. Once the gas distribution is known, we can obtain the total mass density, under the hypothesis of hydrostatic equilibrium. Towards North-East the ICM shows a sudden temperature increase between $0.2$ and $0.4\ {\rm Mpc}$ probably due to a recent merger (see \S~\ref{sec:cold_front}). Such phenomenon does not appear to affect the Southern portion of the cluster, which shows instead a much more regular temperature profile (see panel d) in Fig.~\ref{fig:cf}); we have therefore approximated the overall ICM as isothermal. The total dynamical mass density corresponding to each independent, isothermal gas phase is obtained from:
\begin{displaymath}
\rho_{\rm tot} = -\left( \frac{k_{B}T_{\rm gas}}{4 \pi G \mu m_{\rm p}} \right) \nabla^2 \ln n_{\rm e}
\end{displaymath}
where $k_{B}$ is the Boltzmann constant, $G$ the gravitational constant, $T_{\rm gas}$ the intra-cluster gas temperature and $\mu m_{\rm p}$ the mean particle mass of the gas. For the single and double $\beta$-model, using the cluster average temperature (Table~\ref{tab:spec_cluster}), we obtained (within a sphere with $1\ {\rm Mpc}$ radius) the following values of the total and gas masses:
\begin{eqnarray}
{\rm Global,1\beta}: M_{\rm tot}[1\ {\rm Mpc}]=(3.7\pm0.3)\cdot10^{14}h^{-1}_{72}\ {\rm M}_{\sun}\nonumber\\
M_{\rm gas}[1\ {\rm Mpc}]=(7.4\pm0.5)\cdot10^{13}h^{-5/2}_{72}\ {\rm M}_{\sun}\nonumber\\
{\rm Global,2\beta}: M_{\rm tot}[1\ {\rm Mpc}]=(4.0\pm1.7)\cdot10^{14}h^{-1}_{72}\ {\rm M}_{\sun}\nonumber\\
M_{\rm gas}[1\ {\rm Mpc}]=(7.9\pm1.6)\cdot10^{13}h^{-5/2}_{72}\ {\rm M}_{\sun}\nonumber
\end{eqnarray}
If the assumption of isothermality is dropped and the average cluster temperature profile is used, the resulting cluster total mass is consistent within $10 \%$ with the above values for radii larger than $250\ {\rm kpc}$. At smaller distances to the cluster centre the observed temperature is lower than the cluster average temperature, leading to a systematically lower cluster total mass (within  $25 \%$ of the above values, for radii larger than $75\ {\rm Mpc}$). The use of a temperature profile averaged over $360 \degr$, causes the temperature increase observed in the North-East, which is a spatially localized phenomenon, to be instead associated to the whole cluster, leading nevertheless to systematic errors. The assumption of isothermality is therefore used for the remainder of this paper.\\
Finally, we  introduced the Tail as a separate component in our mass analysis. We modeled the cluster as two dynamically independent systems, with different electron temperatures, seen in projection along the line of sight. These are strict assumptions, but only within this frame can we investigate the effect of the elongated Tail on the mass distribution of the system. Surface brightness profiles were extracted within two sets of annuli shaped according to the results of 2D-fit (see Table~\ref{tab:2Dfit}). The profile for the Northern component was extracted within a $180 \degr$ centered toward the North direction, where the contribute of the Tail is negligible. A corrected 1D profile of the Tail was obtained subtracting the profile of the northern component, taken into account their relative shift and orientation. Best-fit parameters of the resulting surface brightness profile of the Northern component are for a single-$\beta$ model: $S_{0N}=0.32\pm0.03\ [{\rm cts}\ {\rm s}^{-1}\ {\rm arcmin}^{-2}]$, $\beta_{N}=0.389\pm0.007$, $r_{cN}=0.226\pm0.026\ {\rm arcmin}$, $\chi^2_{\rm red.}=4.0$ and for a double-$\beta$ model: $S_{0N1}=0.24\pm0.02\ [{\rm cts}\ {\rm s}^{-1}\ {\rm arcmin}^{-2}]$, $\beta_{N1}=1.3\pm0.3$, $r_{cN1}=0.84\pm0.19\ {\rm arcmin}$, $S_{0N2}=0.029\pm0.004\ [{\rm cts}\ {\rm s}^{-1}\ {\rm arcmin}^{-2}]$, $\beta_{N2}=0.65\pm0.18$, $r_{cN2}=3.1\pm1.1\ {\rm arcmin}$, $\chi^2_{\rm red.}=1.6$. The Tail is instead best described by a single-$\beta$ model: $S_{0S}=0.068\pm0.005\ [{\rm cts}\ {\rm s}^{-1}\ {\rm arcmin}^{-2}]$, $\beta_{S}=1.8\pm1.0$, $r_{cS}=2.8\pm1.0\ {\rm arcmin}$, $\chi^2_{\rm red.}=0.8$. Temperatures of $8.0\ {\rm keV}$ and $0.7\ {\rm keV}$ were used for the Northern and the Southern component, respectively. Under these assumptions, the total mass density was computed adding linearly the mass of the tail to that of the Northern component (each one obtained independently), taking into account their relative positions and orientations. While in the inner regions around the X-ray centroid the mass density is dominated by the Northern component, at large radii the contribution from the tail becomes significant. This is due to both the relative positions of the two components and to the smooth profile of the Tail in contrast to the sharply decreasing one of the Northern component. The resulting values of the total and gas masses are:
\begin{eqnarray}
{\rm N+S},1\ \beta: M_{\rm tot}[1\ {\rm Mpc}]=(4.5\pm1.1)\cdot10^{14}\ {\rm h}^{-1}_{72}\ {\rm M}_{\sun}\nonumber\\
M_{\rm gas}[1\ {\rm Mpc}]=(8.4\pm2.6)\cdot10^{13}h^{-5/2}_{72}\ {\rm M}_{\sun}\nonumber
\end{eqnarray}
Consistent values of the masses are obtained using a double-$\beta$ model for the Northern component. Henceforth with {\textsf N+S} we will imply {\textsf N+S}, $1\ \beta$ model.\\
Although these mass estimates are highly model-dependent, some general qualitative trends emerge. When we assume that all the emission comes from a single structure ({\textsf Global} model), the results of the mass obtained with a single or a double $\beta$-model are similar, especially at small radii (Fig.~\ref{fig:massa}); this reflects the fact that in the central region of the cluster a simple 1-$\beta$ model still provides an acceptable approximation to the surface brightness profile.\\
If we assume instead that the system is a superposition of two distinct systems, we obtain systematically higher values of the mass, especially in the inner regions. This quite general result is a consequence of the non-linear dependence of X-ray luminosity on cluster masses. As seen for the {\textsf Global} model, also for the Northern component both single and double $\beta$-models provide a good description of the surface brightness profile leading to similar final values of the masses.\\
We compute the gas mass fraction at r$_{500}$; to ease comparison with previous work, we quote results for $H_0=50\ \mathrm{km\ s}^{-1} \mathrm{Mpc}^{-1}$ ($\Omega_m=0.3$, $\Omega_{\Lambda}=0.7$). We remark that going out to $r_{500}$ needs a radial extrapolation of $\sim 30\%$. For the three models described above we find: f$_{\rm gas}^{\rm Global,1\beta}(r_{500})=0.33\pm0.04$, f$_{\rm gas}^{\rm Global,2\beta}(r_{500})=0.31\pm0.09$, f$_{\rm gas}^{\rm N+S}(r_{500})=0.39\pm0.14$.  The values of $r_{500}$ for the three models are $1363\ {\rm kpc}$, $1499\ {\rm kpc}$ and $1474\ {\rm kpc}$, respectively. Considerably lower values of f$_{\rm gas}(r_{500})\sim 0.15\pm0.05$ are estimated for relaxed clusters by~\cite{Ett03}.


\section{Comparison of X-ray and  Strong- and Weak-Lensing Mass Estimates}
\label{sec:lensing}
From the results of a spectroscopic survey of faint lensed galaxies in the core of AC~114, \cite{Cam01} revised the lensing model of the cluster derived by~\cite{Nat98} which uses both strong and weak lensing constraints. The revised mass reconstruction made by~\cite{Cam01} includes the strong constraints determined from the redshifts of two multiple-image systems: the 3-image S (${\rm z}=1.86$), that had been already included in the model by~\cite{Nat98}, and the newly discovered 5-image E (${\rm z}=3.347$). \cite{Cam01} found that a mass distribution composed of an elliptical cluster-scale component (the central cluster) plus an additional bimodal component, with the two clumps centered on two other galaxy concentrations, reproduces the observed lensed images. The locations of these two hypothetical additional clumps of matter are shown in the top panel in Fig.~\ref{fig:smoo_energy_bands}. While we detect no excess X-ray emission above the background from Clump 2, diffuse X-ray emission can definitively be seen at Clump 1. It is particularly interesting to note that Clump 1 is located at the Southern edge of the elongated soft Tail.\\

\subsection{Comparing Mass Estimates}
\label{sec:compare_mass}
\begin{figure}
\centering\plotone{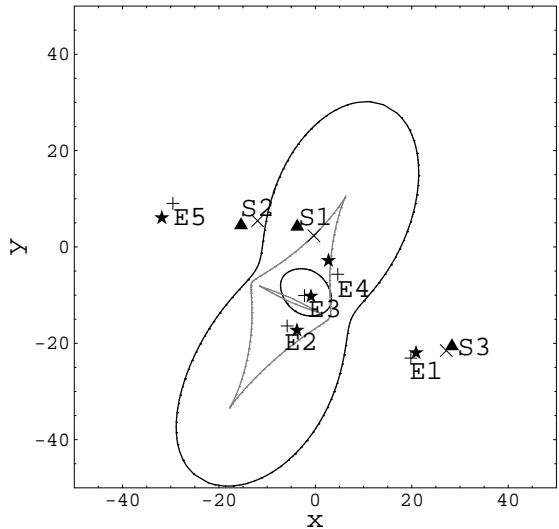}
\caption{Lensing properties of the \textsf{N+S} model, for the two sources at $z=3.347$ (5-images E system) and $z=1.86$ (3-images S system). The black and gray lines represent critical curves and caustics, respectively. Crosses show the image positions of the E-system, X the ones of the S-system; filled stars and triangles the corresponding image positions predicted by the {\textsf N+S} model. The $x$-axis coincides with the North. Distances are in units of arcseconds. The origin coincides with the position of the X-ray centroid.}
\label{fig:lensing}
\end{figure}

From their detailed analysis of weak and strong lensing data for AC~114,~\cite{Nat98} estimated the cluster mass within  $75.0$, $150.0$ and $500.0$ kpc of the cluster center. Their results are listed in Table~\ref{tab:lens_mass} together with our values for the projected X-ray masses.\\
The X-ray and lensing masses are in  good agreement, for the two-component X-ray models, even at the smallest radii. In particular, ratios ${\rm M}_{\rm lens}/{\rm M}_{\rm X}$ at 75 kpc are $1.23 \pm 0.16$, $1.1 \pm 0.5$, and $0.95 \pm 0.16$ for the \textsf{Global}-1$\beta$, \textsf{Global}-2$\beta$, and \textsf{N+S} models, respectively. The results are remarkable in view of results from some other unrelaxed clusters in which  the X-ray-determined masses are only one-third to one-half the lensing masses at radii less than 100 kpc~\citep{Mac02,Xue02}.\\
The use of a detailed map of the surface mass density is crucial to accurate modeling of strong lensing phenomena. \cite{Cam01} used the galaxy density to predict a cluster mass distribution that reproduced the observed strong-lensing images.\\
Since the X-ray emission supplies an independent tracer of the underlying cluster mass, we used our inferred  (X-ray-derived) mass distribution for a new strong lensing analysis.  For this purpose we used the {\sf gravlens} software\footnote{\sf http://astro.uchicago.edu/\~ckeeton/gravlens}~\citep{Kee01}. We found that the mass distribution computed from the \textsf{Global} model is sub-critical, and is therefore unable to produce any strong lensing. The \textsf{N+S} model produces two lips-shaped caustics, one inside the other, oriented perpendicular to one another~\citep{Schn92} as shown in Fig.~\ref{fig:lensing}. Sources inside the inner caustic are multiple imaged in patterns similar to the ones observed in the E-systems. A better match with the lensed images is obtained when the center of the Northern component is slightly shifted toward the direction of the cD galaxy; this provides a reasonably good reproduction of both E and S systems with a resulting average distance between the lensed images and their best-fit positions: $d_{E}=2.4\arcsec$ for the five images in the E-system and $d_{S}=2.9\arcsec$ for the three images in the S-system (see Fig.~\ref{fig:lensing}). A detailed new strong lensing analysis of AC~114, based on the X-ray mass results obtained in this paper is the object of a forthcoming paper~\citep{Ser04}.\\

\begin{deluxetable}{ccccc}
\tablecolumns{8}
\tablewidth{0pc}
\tablecaption{Lensing and X-ray mass measurements}
\tablehead{
\colhead{Radius} & \colhead{M$_{\rm X}$} & \colhead{M$_{\rm X}$} & \colhead{M$_{\rm X}$} & \colhead{M$_{lens}$} \\
\colhead{[kpc]}  &	\colhead{Global 1-$\beta$} &	\colhead{Global 2-$\beta$}   &\colhead{N$+$S} &\colhead{}
}
\startdata
75.0  & $3.4 \pm 0.3 $  &  $ 3.7  \pm 1.5$ & $4.4 \pm 0.6$  &  $4.2 \pm 0.4$ \\
150.0 & $9.7 \pm 0.7 $  &  $ 10   \pm 4$   & $11.2 \pm 1.7$ &  $12  \pm 1.5$ \\
500.0 & $39  \pm 3   $  &  $ 41   \pm 16$  & $47 \pm 10$    &  $40  \pm 4  $ \\
\enddata
\tablecomments{Column 1: radius. Cols. 2-4: Projected X-ray masses. Column 5: $M_{\rm lens}$ from~\cite{Nat98}. All results in this Table are given for: $H_0=50\ \mathrm{km\ s}^{-1} \mathrm{Mpc}^{-1}$. All values of the masses are given in $[10^{13}\ {\rm M}_{\sun}]$.}
\label{tab:lens_mass}
\end{deluxetable}


\section{Summary and Conclusions}
We have presented the first Chandra observation of the galaxy cluster AC~114, which has revealed a strongly irregular X-ray morphology. Below $0.5\ {\rm keV}$ the X-ray emission is dominated by two main components: the cluster, roughly centered on the optical position of AC~114, and a diffuse `` Tail'', extending almost $400$ kpc, from the cluster center to the south-east. The Tail has an energy distribution different from the rest of the cluster, showing significant excess of soft emission. The Tail connects the cluster core with the location of Clump 1, an additional clump of matter previously inferred from lensing analysis. We also detect diffuse X-ray emission from Clump 1, confirming its existence. We conclude that Clump 1 is probably associated with the  cluster, and may have interacted with it.  \\
We propose a possible scenario in which Clump 1, located nearby a large over-density of galaxies, in its motion from the north-west through the cluster, has been ram-pressure stripped of most of its intra-group gas, now still visible as the soft Southern Tail. During its interaction with the cluster, Clump 1 would have also distorted the intra-cluster gas, causing the asymmetrical stretch of the cluster emission toward south-east. Such a picture is supported by recent adaptive mesh hydro/N-body simulations~\citep{Bur03} which show soft extended features similar to our tail forming after substructure mergers.\\
AC~114 also shows further signs of recent dynamical activity close to the cluster center. A sharp decrease in the cluster surface brightness toward north-east and a corresponding peak in the gas temperature are observed, suggesting the motion of a sub-structure toward this direction which compresses and heats intra-cluster gas ahead of its path. \\
These two putative interactions take place along different directions in the plane of the sky. Their time-scales are also different. High-temperature fronts forming in front of fast moving sub-structures are short-living phenomena, and are therefore signs of ``recent'' merging processes; comparison with simulations by~\cite{Mat03} lead to an estimated time of the merger of $\approx 200\ {\rm Myr}$. In the Tail region there is instead no sign of high velocities involved, and considering the length of the Tail, the closest approach of Clump 1 to the core must have been a much earlier event ($\approx 800\ {\rm Myr}$ ago if one assumes the velocity of Clump 1 to be $\approx 500\ {\rm km\ s}^{-1}$). These two merger events were thus likely to have been independent phenomena. We note that in a hierarchical cosmological scenario, in which clusters form from accretion of matter through sets of intersecting filaments, multiple merging processes, directed along different axes, are expected.\\
In spite of this evidence for complex interactions, we have obtained estimates of the X-ray masses that are in remarkably good agreement with lensing results, even at distances as small as 75 kpc from the cluster center. The simplest isothermal model for the X-ray emission is under-critical and thus cannot reproduce the lensing results. Two-component models, in which the components interact, or are seen in projection along the line of sight, however, provide a consistent estimate of the total mass and approximately reproduce the positions of strongly lensed background sources, though the best agreement is found if the X-ray emitting gas is displaced slightly with respect to the potential minimum. Agreement between X-ray and lensing mass estimates has recently been found also in another irregular cluster (MS1008.1-1224) by~\cite{Ett03b}. We therefore advocate for the robustness of the X-ray mass estimates, and conclude that the assumption of hydrostatic equilibrium can yield accurate mass estimates even in clusters as dynamically active as AC~114 once the gas density distribution is properly mapped.


\acknowledgements
\noindent Al mio pap\'a\\
A special thanks goes to C.~R. Keeton for providing the software for the lensing analysis and helping us to use it at its best, and to John Arabadjis for his continuous help during the past year. This work has been supported by NASA grants NAS8-39073 and NAS8-00128. We have extensively used the NASA/IPAC Extragalactic Database (NED) which is operated by the Jet Propulsion Laboratory, California Institute of Technology, under contract with the National Aeronautics and Space Administration.

\end{document}